\documentclass{article}
\usepackage{frascatiphys,here,graphicx,subfigure}
\begin{document}
\title{ 
DOUBLE FLAVOR VIOLATING TOP QUARK DECAYS\\
IN EFFECTIVE THEORIES
}
\author{
Carmine Pagliarone\\
{\em Universit\'a di Cassino \& Istituto Nazionale di Fisica Nucleare Pisa, Italy.} \\
A. Fern\' andez and J. J. Toscano        \\
{\em Benem\'erita Universidad Aut\'onoma de Puebla, Puebla, Pue., M\'exico.}
}
\maketitle
\baselineskip=9.pt
%%%%%%%%%%%%%%%%%%%%%%%%%%%%%%%%%%%%%%%%%%%%%%%%%%%%%%%%%%%%%%%%%%%%%%%%%%
\begin{abstract}
The possibility of detecting double flavor violating top quark transitions
$t \to u_i\tau \mu$ ($u_i=u,c$) is explored in a model--independent manner, 
using the effective Lagrangian approach.
Low--energy data, on high precision measurements, and current
experimental limits are used to constraint the $tu_iH$ and $H\tau \mu$ 
vertices and then to calculate the branching ratio BR$(t \to u_i\tau \mu)$. 
If in the Standard Model BR$(t \to u_i\tau \mu)$ is of the order of $10^{-13}$$-10^{-14}$, 
higgs--mediated double flavor violating top quark decays can occur with branching 
ratios ranging from $10^{-3}$ to $10^{-4}$ for $114.4$ GeV$/c^2$ $< m_H<$ $2m_W$,
that is at the reach of the CERN Large Hadron Collider.
\end{abstract}
%%%%%%%%%%%%%%%%%%%%%%%%%%%%%%%%%%%%%%%%%%%%%%%%%%%%%%%%%%%%%%%%%%%%%%%%%%
%%
%%
%%
\baselineskip=11pt
%%
%%%%%%%%%%%%%%%%%%%%%%%%%%%%%%%%%%%%%%%%%%%%%%%%%%%%%%%%%%%%%%%%%%%%%%%%%%
\section{Introduction}
%%%%%%%%%%%%%%%%%%%%%%%%%%%%%%%%%%%%%%%%%%%%%%%%%%%%%%%%%%%%%%%%%%%%%%%%%%
Despite of the fact that the top quark is the heaviest known
particle, with a mass comparable to the electroweak symmetry
breaking scale, its dynamical behavior is rather restrictive. 
Within the Standard Model (SM), the top quark production cross section is 
evaluated with an uncertainty that is of the order of $\sim 15\%$ and top quarks
are assumed to decay to a $W$ boson and a $b$ quark almost $100\%$ of the time.
%%%%
Due to its exceedingly heavy mass, it is reasonable to expect the top to
be more related to new physics than other fermions.
%%%%%
Top quark physics may then serve as an important window for probing physics beyond
the SM. 
%%%%
Although some properties of the top quark have
already been examined at the Tevatron\cite{TopTevatron}, a
further scrutiny is expected at the CERN Large Hadron Collider
(LHC), which will operate as a veritable top quark factory,
producing about eight millions of $\bar{t}t$ events per year in
its first stage, and hopeful up to about eighteen millions in
subsequent years\cite{TopLHC}. Other relevant studies will be
realized in the next generation $e^+e^-$ linear colliders.
%%
%%%%%%%%%%%%%%%%%%%%%%%%%%%%%%%%%%%%%%%%%%%%%%%%%%%%%%%%%%%%FIGURE 1
\begin{figure}
\centering
\includegraphics[width=2.2in]{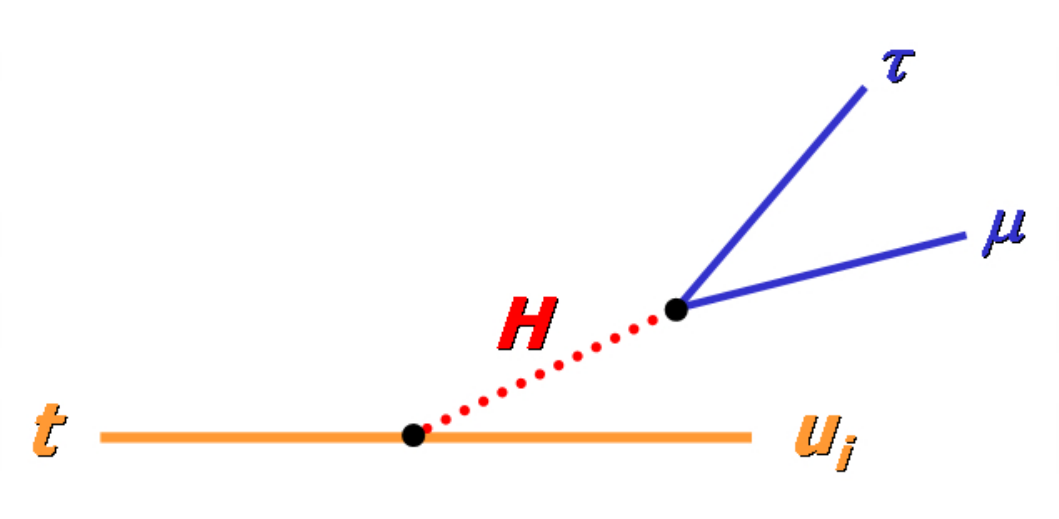}
\caption{\label{TD} Diagram for double flavor violating (DFV) top quark
decay.}
\vspace{-.35cm}
\end{figure}
%%%%%%%%%%%%%%%%%%%%%%%%%%%%%%%%%%%%%%%%%%%%%%%%%%%%%%%%%%%%%%%%%%%

%%
%%
%%%%%%%%%%%%%%%%%%%%%%%%%%%%%%%%%%%%%%%%%%%%%%%%%%%%%%%%%%%%%%%%%%%
\section{Effective Lagrangian for the Yukawa sector}
%%%%%%%%%%%%%%%%%%%%%%%%%%%%%%%%%%%%%%%%%%%%%%%%%%%%%%%%%%%%%%%%%%%
\label{y} In the SM, the Yukawa sector is both CP and flavor 
conserving. CP and flavor violation can be generated at the tree
level if new scalar fields are introduced. Another alternative, which
does not contemplate the introduction of new degrees of freedom,
consists in incorporating into the classical action the virtual
effects of the heavy degrees by introducing $SU_L(2)\times
U_Y(1)$--invariant operators of dimension higher than four.
A Yukawa sector with these features has the following structure\cite{T2,T3}
\begin{eqnarray}
{\cal L}^Y_{eff}&=&-Y^l_{ij}(\bar{L}_i\Phi
l_j)-\frac{\alpha^i_{ij}}{\Lambda^2}(\Phi^\dag \Phi)(\bar{L}_i\Phi
l_j)+ H.c. \nonumber \\
&&-Y^d_{ij}(\bar{Q}_i\Phi
d_j)-\frac{\alpha^d_{ij}}{\Lambda^2}(\Phi^\dag \Phi)(\bar{Q}_i\Phi
d_j)+ H.c.\\
&&-Y^u_{ij}(\bar{Q}_i\tilde{\Phi}
u_j)-\frac{\alpha^u_{ij}}{\Lambda^2}(\Phi^\dag
\Phi)(\bar{Q}_i\tilde{\Phi} u_j)+ H.c.\nonumber
\end{eqnarray}
where $Y_{ij}$, $L_i$, $Q_i$, $\Phi$, $l_i$, $d_i$, and $u_i$
stand for the usual components of the Yukawa matrix, the
left--handed lepton doublet, the left--handed quark doublet, the
Higgs doublet, the right--handed charged lepton singlet, and the
right--handed quark singlets of down and up type, respectively.
The $\alpha_{ij}$ numbers are the components of a $3\times 3$
general matrix, which parameterize the details of the underlying
physics, whereas $\Lambda$ is the typical scale of these new
physics effects.
After spontaneous symmetry breaking, this extended Yukawa sector
can be diagonalized as usual via the unitary matrices
$V^{l,d,u}_L$ and $V^{l,d,u}_R$, which relate gauge states to mass
eigenstates. In the unitary gauge, the diagonalized Lagrangian can
be written as follows:
%\begin{eqnarray}
%{\cal
%L}^Y_{eff}&=&-\Big(1+\frac{g}{2m_W}H\Big)\Big(\bar{E}M_lE+\bar{D}M_dD+\bar{U}M_uU\Big)\nonumber\\
%&&-H\Big(1+\frac{g}{4m_W}H\Big(3+\frac{g}{2m_W}H\Big)\Big)\Big(\bar{E}\Omega^lP_RE+\bar{D}\Omega^d
%P_RD+\bar{U}\Omega^u P_RU+H.c.\Big)
%\end{eqnarray}
%%
\begin{eqnarray}
{\cal
L}^Y_{eff}=-\Big(1+\frac{g}{2m_W}H\Big)\Big(\bar{E}M_lE+\bar{D}M_dD+\bar{U}M_uU\Big)\nonumber\\
-H\Big(1+\frac{g}{4m_W}H\Big(3+\frac{g}{2m_W}H\Big)\Big)\Big  
(\bar{E}\Omega^lP_RE+\bar{D}\Omega^d
P_RD+\bar{U}\Omega^u P_RU+H.c.\Big) 
\end{eqnarray}
where the $M_a$ ($a=l,d,u$) are the diagonal mass matrix and
$\bar{E}=(\bar{e},\bar{\mu},\bar{\tau})$,
$\bar{D}=(\bar{d},\bar{s},\bar{b})$, and
$\bar{U}=(\bar{u},\bar{c},\bar{t})$ are vectors in the flavor
space. In addition, $\Omega^a$ are matrices defined in the flavor
space given by
\begin{equation}
\Omega^a=\frac{1}{\sqrt{2}}\Big(\frac{v}{\Lambda}\Big)^2V^a_L\alpha^aV^{a\dag}_R
\end{equation}
The above effective Lagrangian describes the most general coupling of
renormalizable type of a scalar field to pairs of fermions. This 
reproduces the main features of most of extended Yukawa sectors,
as the most general version of the two--Higgs doublet model
(THDM-III)\cite{THDM-III} and  multi--Higgs models that comprise
additional multiplets of $SU_L(2)\times U_Y(1)$ or scalar
representations of larger gauge groups. Our approach also cover
more exotic formulations of flavor violation, as the so--called
familons models\cite{Familons} or theories that involves an
Abelian flavor symmetry\cite{AFS}. Results, presented in this paper, 
are then applicable to a wide variety of models that predict 
scalar--mediated flavor violation.
%%%%%%%%%%%%%%%%%%%%%%%%%%%%%%%%%%%%%%%%%%%%%%%%%%%%%%%%%%%%%%%%%%%%%%%%%%%

%%%%%%%%%%%%%%%%%%%%%%%%%%%%%%%%%%%%%%%%%%%%%%%%%%%%%%%%%%%%%%FIGURE 4-5
\begin{figure}[t!H]
    \begin{center}
%\subfigure[]
        {\includegraphics[scale=0.34]{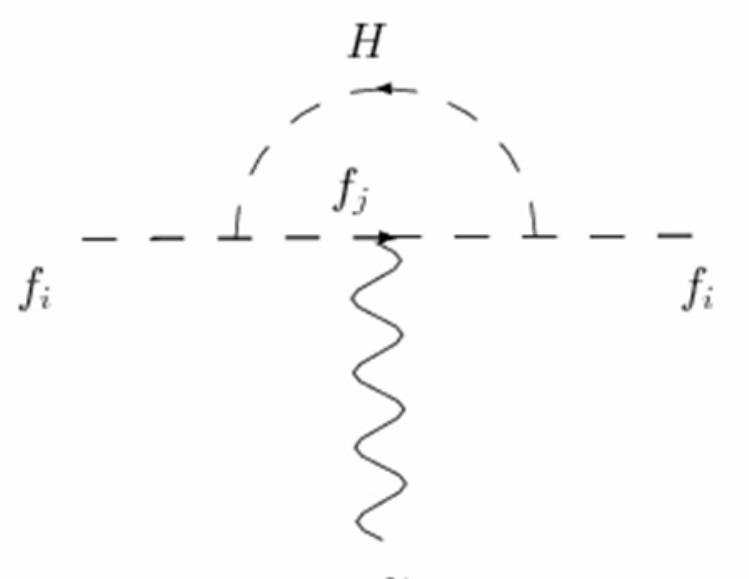}}
%\subfigure[]
        {\hspace{1.5cm}\includegraphics[scale=0.34]{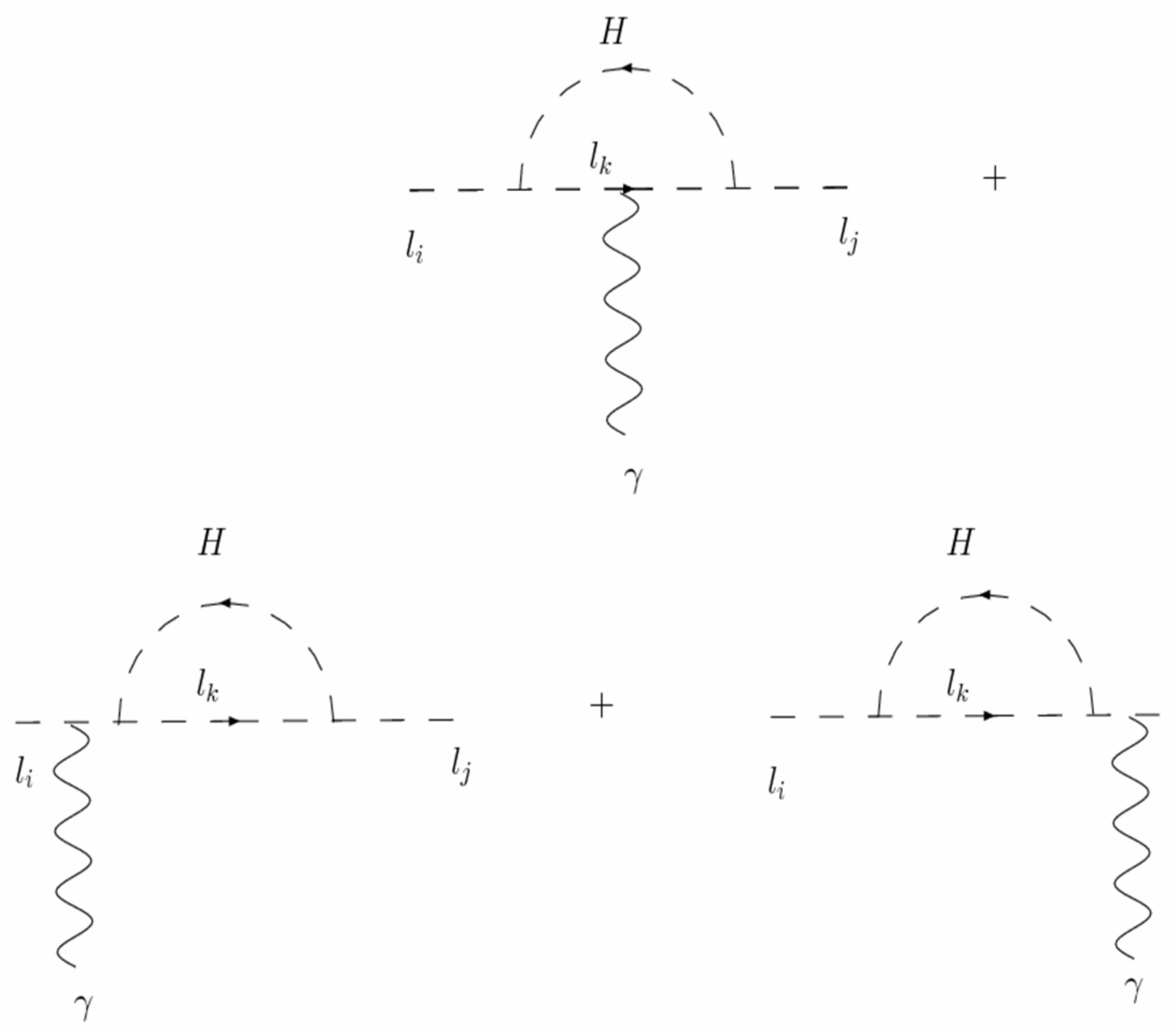}}
        \caption{\it a) \label{BT} Diagram contributing to the magnetic and
electric dipole moments of the $f_i$ fermion; b) \label{MD} diagrams contributing 
to the $l_i \to l_j\gamma$ decay.}
%% Figure 3 & 4
\label{exfig}
    \end{center}
\vspace{-0.75cm}
\end{figure}
%%%%%%%%%%%%%%%%%%%%%%%%%%%%%%%%%%%%%%%%%%%%%%%%%%%%%%%%%%%%%%%%%%%%%%%%%%%%%%%%%%%%%%%%%

%%%%%%%%%%%%%%%%%%%%%%%%%%%%%%%%%%%%%%%%%%%%%%%%%%%%%%%%%%%%%%%%%%%%%%%%%%% THE DECAY top...
\section{The decay $t\to u_i\tau \mu$}
%%%%%%%%%%%%%%%%%%%%%%%%%%%%%%%%%%%%%%%%%%%%%%%%%%%%%%%%%%%%%%%%%%%%%%%%%%%
\label{d} The branching ratio for the $3-$body decay $t\to u_i\tau \mu$ is 
calculated using the general couplings $t\to u_iH$ and $H\to \tau \mu$, 
obtained using the Effective Lagrangian technique (for further details see\cite{papunderprep}):
\begin{eqnarray}
Br(t\to
u_iH)&=&\frac{\Omega^2_{tu}}{32\pi}\Bigg(\frac{m_t}{\Gamma_t}\Bigg)\left(1-\Big(\frac{m_H}{m_t}\Big)^2\right)^2\\
Br(H\to \tau \mu)&=&\frac{\Omega^2_{\tau
\mu}}{4\pi}\Bigg(\frac{m_H}{\Gamma_H}\Bigg)\left(1-\Big(\frac{m_\tau}{m_H}\Big)^2\right)^2
\end{eqnarray}
These two expressions have been obtained ignoring the $u_i$ and the muon
masses. Now, in order to evaluate the branching ratio
Br$(t\to u_i\tau \mu)=$ Br$(t\to u_iH)$ $\times$ Br$(H\to \tau \mu)$ we need to calculate  
the following two terms: $|\Omega_{u_{i}t}|^2$ and $|\Omega_{\tau\mu}|^2$.

%%%%%%%%%%%%%%%%%%%%%%%%%%%%%%%%%%%%%%%%%%%%%%%%%%%%%%%%%%%%%%%%%%%%%%%%%%% THE DECAY top...
\section{Bounding the $tu_iH$ and $H\tau \mu$ vertices}
%%%%%%%%%%%%%%%%%%%%%%%%%%%%%%%%%%%%%%%%%%%%%%%%%%%%%%%%%%%%%%%%%%%%%%%%%%%
An estimate of the size of $|\Omega_{u_{i}t}|^2$ and $|\Omega_{\tau\mu}|^2$
parameters is needed. Although they have been already estimated before, by
means of the Cheng--Sher ansazt\cite{SHER1} or with other similar
assumptions\cite{T2}, we will resort to the available low--energy data,
in order to be more independent from any extra assumption.
A constrain on the $|\Omega_{u_{i}t}|^2$
parameter can be obtained, from the low--energy data, through one--loop or higher
order effects. We will examine the contribution to the $t u_{i} H$ vertex to the 
proton and neutron magnetic and electric dipole moments (see Figure~\ref{BT}.a).
The contribution of the $f_{i} f_{j}H$ vertex to the magnetic and electric dipole 
moments of the $f_{i}$ fermion have been calculated and the electromagnetic
dipoles of $f_i$ can be written as follows:
\begin{equation}
a_i=-\frac{Q_jx_ix_j}{8\pi^2}Re(\Omega^2_{ij})g(x_j), \;\;\;\;\;\;\;\;\;
d_i=-\frac{Q_jx_j}{16\pi^2}\Big(\frac{e}{m_H}\Big)Im(\Omega^2_{ij})g(x_j)
\end{equation}
%%
%%%%%%%%%%%%%%%%%
%%
\noindent
We now are in position of using low--energy data to bound the
$|\Omega_{tu_{i}}|^2$ and $|\Omega_{\tau \mu}|^2$ parameters.

%%%%%%%%%%%%%%%%%%%%%%%%%%%%%%%%%%%%%%%%%%%%%%%%%%%%%%%%%%%%FIGURE 2-3
\begin{figure}[t!H]
    \begin{center}
%\subfigure[]
        {\includegraphics[scale=0.18]{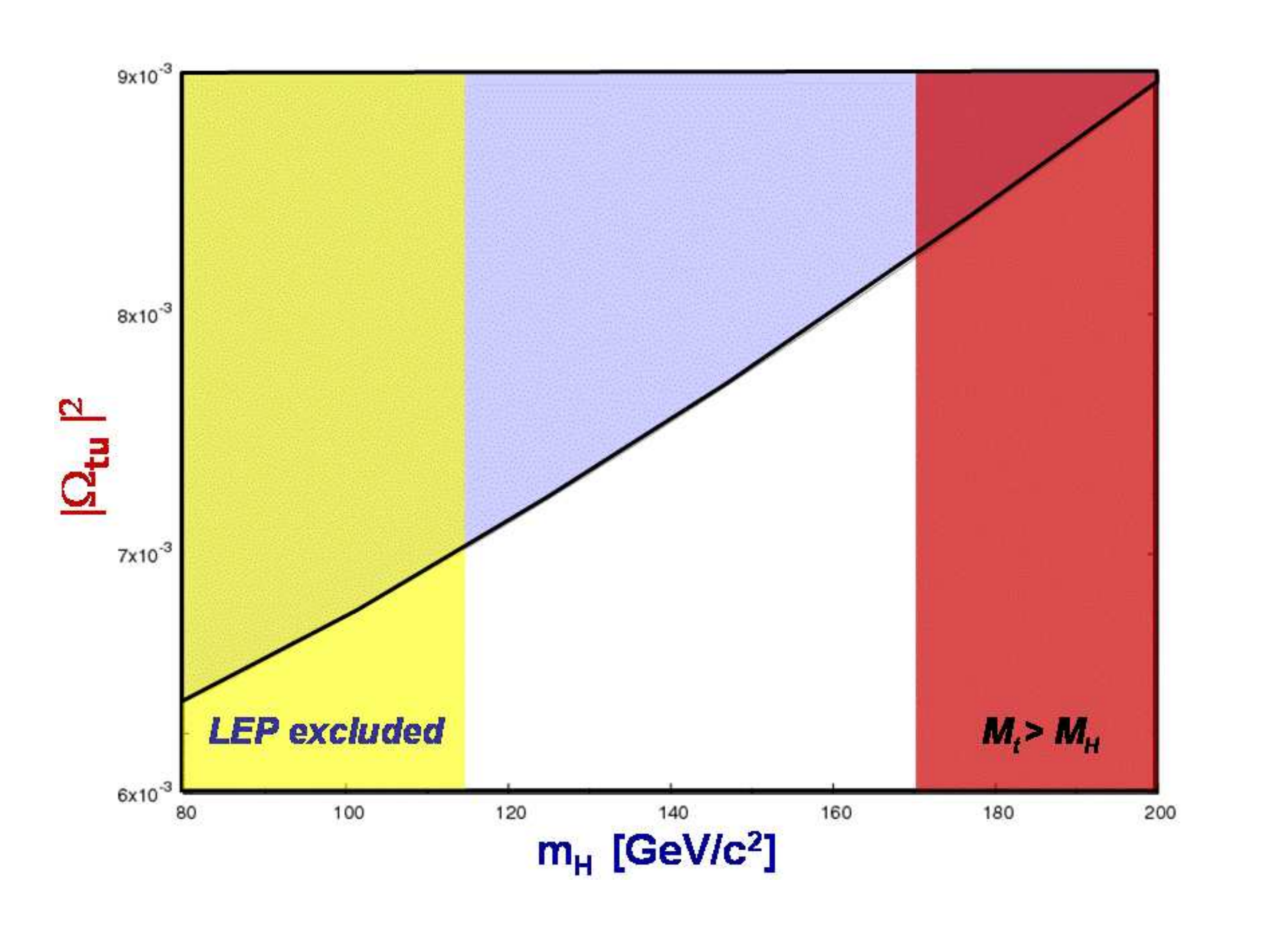}}
%\subfigure[]
        {\includegraphics[scale=0.17]{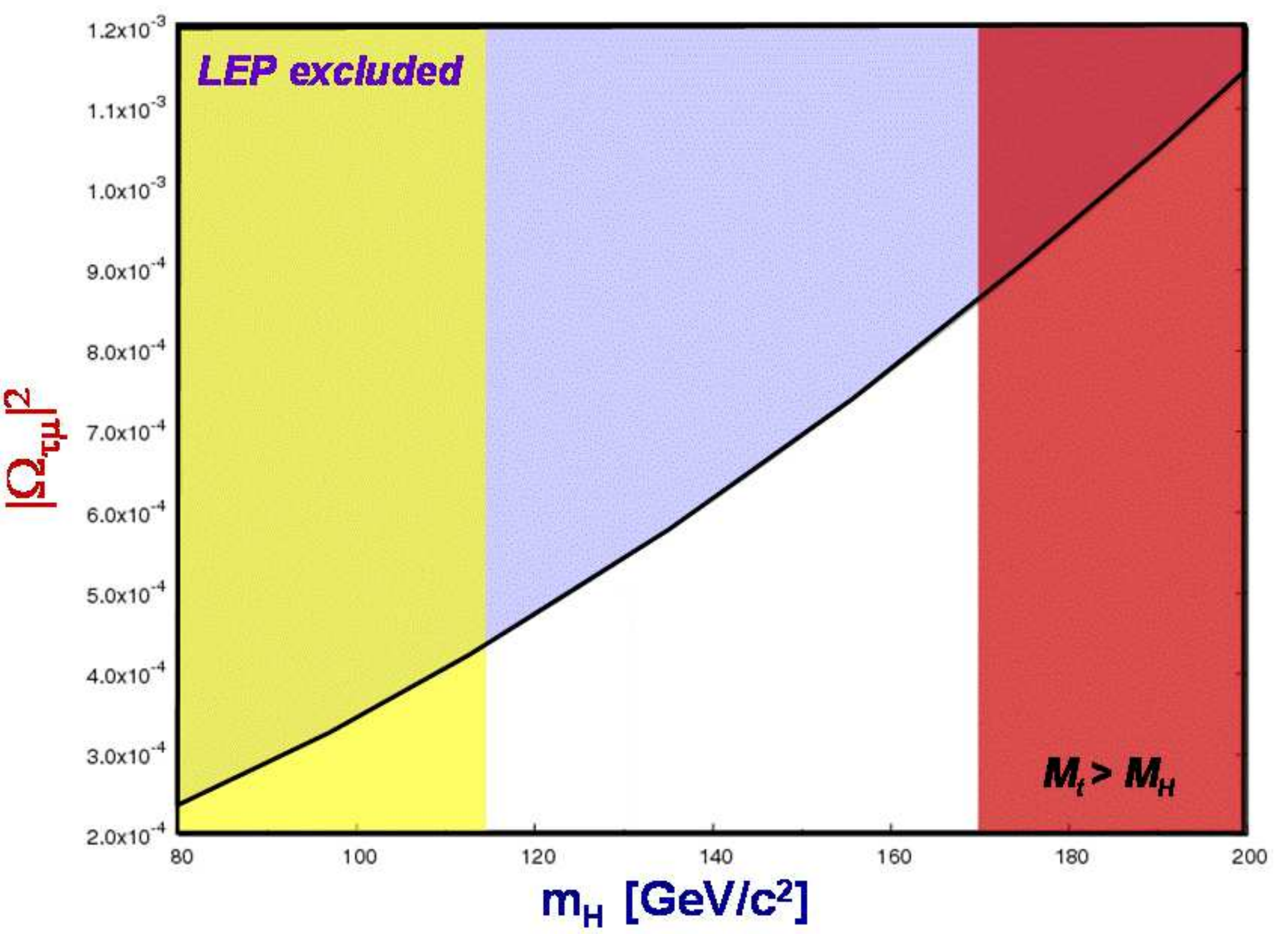}}
        \caption{\it a) \label{otu}Behavior of $\Omega^2_{tu}$ b) $\Omega^2_{\tau \mu}$, as a function of the Higgs mass.}
%% Figure 2 & 3
\label{exfig}
    \end{center}
\vspace{-0.8cm}
\end{figure}
%%%%%%%%%%%%%%%%%%%%%%%%%%%%%%%%%%%%%%%%%%%%%%%%%%%%%%%%%%%%%%%%%%%%%%%%%%%%%%%%%%%%%%%%%

%%%%%%%%%%%%%%%%%%%%%%%%%%%%%%%%%%%%%%%%%%%%%%%%%%%%%%%%%%%%%%%%%%%%%%%%%%%
\subsection{Bounds on $|\Omega_{tu_{i}}|^2$}
%%%%%%%%%%%%%%%%%%%%%%%%%%%%%%%%%%%%%%%%%%%%%%%%%%%%%%%%%%%%%%%%%%%%%%%%%%%
In order to bound $Re(\Omega^2_{tu})$ we used the available precision
measurements on the proton and neutron magnetic dipole moments. We used 
the experimental uncertainty on the proton magnetic dipole, as it is the 
most restrictive. Thus, the contribution of the $tuH$ vertex must be less 
than this uncertainty:\cite{PDG}
\begin{equation}
|\Delta a^{Exp}_p|< 2.8 \times 10^{-8}
\end{equation}
In order to constrain the $Im(\Omega^2_{tu})$ term, we used the very stringent 
experimental limit coming from the neutron electric dipole moment\cite{EDMN}:
\begin{equation}
|d_n|<2.9\times 10^{-26}\ e.cm.
\end{equation}
The $p$ and $n$ magnetic and electric dipole moments, respectively, 
are related with those of their elementary constituents through the 
following expressions:
\begin{equation}
a_p=\frac{2}{3}a_u-\frac{1}{3}a_d-\frac{1}{3}a_s,  \;\;\;\;\;\;\;\;\;\;\;\;\;\;\;
d_n=\frac{4}{3}d_d-\frac{1}{3}d_u
\end{equation}
Assuming that the new physics contributions to $a_p$ and $d_n$ arise
exclusively from the quark $u$, as in this case $f_j=t$
and assuming, as usual, $m_d\approx m_u\approx m_n/3\approx
m_p/3$ we finally get:
\begin{equation}
|Re(\Omega^2_{tu})|<(54\pi^2)\Bigg|\frac{\Delta
a^{Exp}_p}{x_tx_pg(x_t)}\Bigg|, \;\;\;\;\;\;\;
|Im(\Omega^2_{tu})|<(36\pi^2)\Big(\frac{m_H}{e}\Big)\Bigg|\frac{d^{Exp}_n}{x_tg(x_t)}\Bigg|
\end{equation}
%%
%%%%%%%%%%%%%%%%%%%%%%%%%%%%%%%%%%%%%%%%%%%%%%%%%%%%%%%%%%%%%%%%%%%%%%%%%%%%%%%%%%%%%% >>>>>>>>>> from here...
As in this case $|Im(\Omega^2_{tu})<<$ $|Re(\Omega^2_{tu})|$ (as matter of fact for $m_H=100$ GeV$/c^2$ we have $|Re(\Omega^2_{tu})|< 6\times 10^{-3}$ and 
$|Im(\Omega^2_{tu})|< 10^{-7}$) we can write:
\begin{equation}
|\Omega_{tu}|^2<(54\pi^2)\Bigg|\frac{\Delta
a^{Exp}_p}{x_tx_pg(x_t)}\Bigg|
\end{equation}
In Figure~\ref{otu}.a, we show the behavior of $|\Omega_{tu}|^2$
as a function of the Higgs mass.

%%%%%%%%%%%%%%%%%%%%%%%%%%%%%%%%%%%%%%%%%%%%%%%%%%%%%%%%%%%%%%%%%%%%%%%%%%%
\begin{figure}
\centering
\includegraphics[width=2.8in]{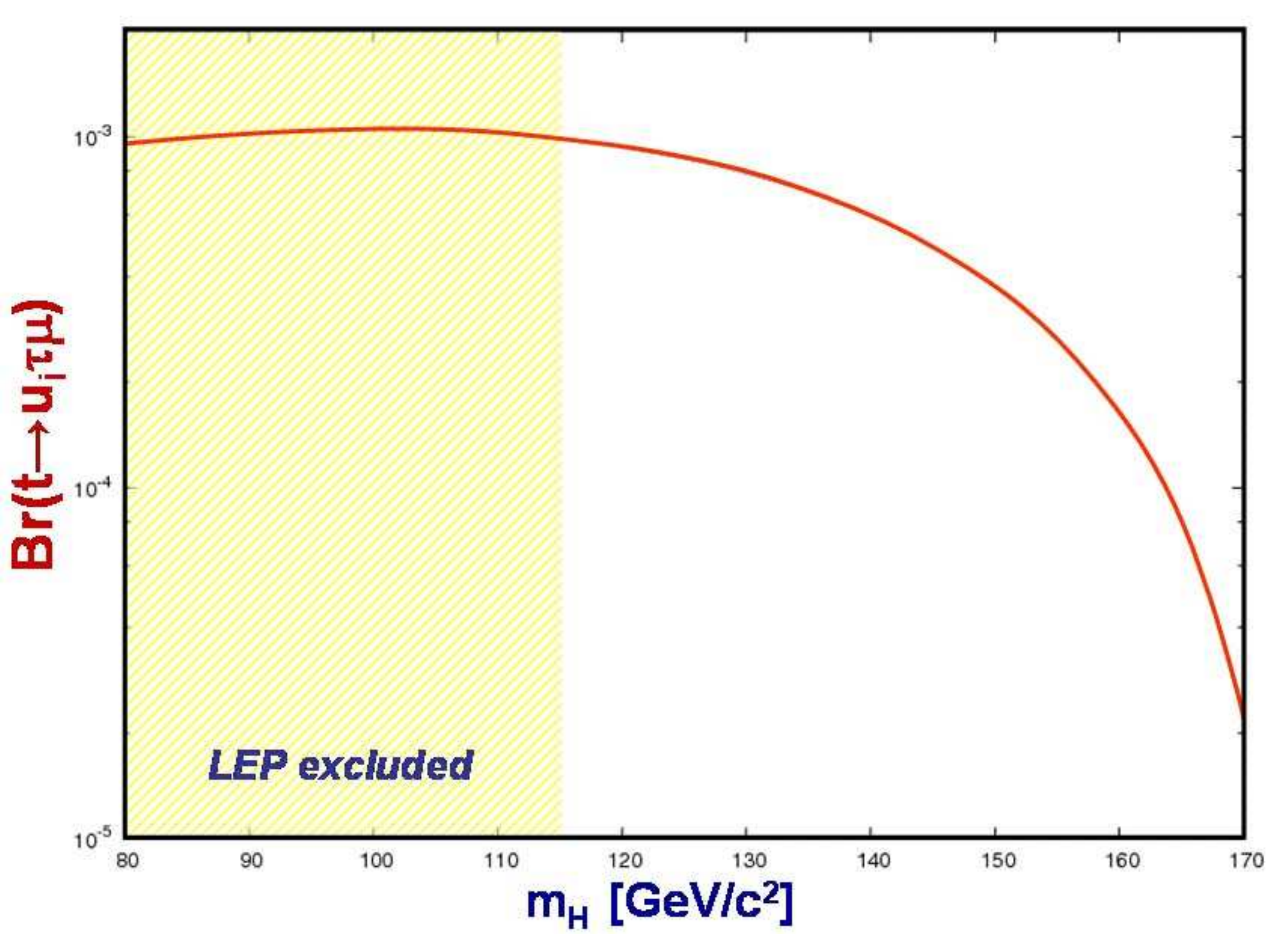}
\caption{\label{BR}Branching ratio for the $t\to u_i\tau \mu$
decay as a function of the Higgs mass.}
\vspace{-0.4cm}
\end{figure}
%%%%%%%%%%%%%%%%%%%%%%%%%%%%%%%%%%%%%%%%%%%%%%%%%%%%%%%%%%%%%%%%%%%%%%%%%%%

%%%%%%%%%%%%%%%%%%%%%%%%%%%%%%%%%%%%%%%%%%%%%%%%%%%%%%%%%%%%%%%%%%%%%%%%%%%
\subsection{Bounds on $|\Omega_{\tau \mu}|^2$}
%%%%%%%%%%%%%%%%%%%%%%%%%%%%%%%%%%%%%%%%%%%%%%%%%%%%%%%%%%%%%%%%%%%%%%%%%%%
The best experimental limits on the $\mu$ anomalous magnetic moment, and on 
the electric dipole moment are\cite{PDG}:
\begin{equation}
|\Delta a^{Exp}_\mu|<5.4 \times 10^{-10}, \;\;\;\;\;\;\;\;|d^{Exp}_\mu|<3.7\times 10^{-19} \ e.cm.
\end{equation}
In this case we obtain:
\begin{equation}
|Re(\Omega^2_{\tau \mu})|<(8\pi^2)\Bigg|\frac{\Delta
a^{Exp}_\mu}{x_\tau x_\mu g(x_\tau)}\Bigg|, \;\;\;\;\;
|Im(\Omega^2_{\tau
\mu})|<(16\pi^2)\Big(\frac{m_H}{e}\Big)\Bigg|\frac{d^{Exp}_\mu}{x_\tau
g(x_\tau)}\Bigg|
\end{equation}
%%%%%%%%%%%%%%%%%%%%%%%%%%%%%%%%%%%%%%%
This time we have $|Re(\Omega^2_{tu})|<$ $|Im(\Omega^2_{tu})$ 
(as matter of fact for $m_H=100$ GeV$/c^2$ we get 
$|Re(\Omega^2_{\tau \mu})|< 3\times 10^{-4}$ and
$|Im(\Omega^2_{\tau \mu})|< 2$).
%%%%%%%%%%%%%%%%%%%%%%%%%%%%%%%%%%%%%%%
In order to simplify the analysis we will assume that the 
leptonic Yukawa sector is also CP--conserving, then we can write:
\begin{equation}
\Omega^2_{\tau \mu}<(8\pi^2)\Bigg|\frac{\Delta a^{Exp}_\mu}{x_\tau
x_\mu g(x_\tau)}\Bigg|
\end{equation}
In Figure~\ref{otu}.b, we show the behavior of $\Omega^2_{\tau \mu}$ as a
function of the higgs mass.
%%%%%%%%%%%%%%%%%%%%%%%%%%%%%%%%%%%%%%%%%%%%%%%%%%%%%%%%%%%%%%%%%%%%%%%%%%%%%%%
\section{DFV top quark decays at LHC}
%%%%%%%%%%%%%%%%%%%%%%%%%%%%%%%%%%%%%%%%%%%%%%%%%%%%%%%%%%%%%%%%%%%%%%%%%%%%%%%
We are now able to evaluate the branching ratio for the decay $t \to u_i\tau \mu$ 
as a function of the higgs mass: Br$(t\to u_i\tau \mu)=$ Br$(t\to u_iH)$ $\times$ 
Br$(H\to \tau \mu)$. The result, given in Figure~\ref{BR}, shows that 
Br$(t\to u_i\tau \mu)$ ranges from $10^{-3}$ to $10^{-4}$ for a higgs mass between 
$114.4$ GeV$/c^2$ and $2m_W$.\\
%%%
LHC will be the ideal machine for investigating the characteristic of the heaviest quark and its
role in the SM; with an NLO production cross-section of about 830 pb, 2 top-pair per second are
expected and more than 10 million per year at low luminosity conditions of $2\times 10^{33} cm^{-2} s^{-1}$.
%%
%%%%%%%%%%%%%%%%%%%%%%%%%%%%%%%%%%%%%%%%%%%%%%%%%%%%%%%%%%%%%%%%%%%%%%%%%%%%%%%%%%%%%%%%%%%%%%%%%%% 
If $\mathcal{B}$ represents the branching ratio for the higgs mediated double flavor violating top 
quark decay, $\mathcal{B} \equiv$ Br$(t\to u_i\tau \mu)$, and if we assume no other 
significantly accessible decay channels, either than the SM and the DFV, we will have: 
Br$(t \rightarrow W b)= 1 - \mathcal{B}$.
%%%%%%%%%%%%%%%%%%%%%%%%%%%%%%%%%%%%%%%%%%%%%%%%%%%%%%%%%%%%%%%%%%%%%%%%%%%%%%%%%%%%%%%%%%%%%%%%%%% 
%%%
A $t \bar{t}$ pair can then decays as follows: 
a) purely SM decays (SM-SM), where both the top quarks decay into $Wb$ (Br$(t \bar t)|_{SM}=(1- \mathcal{B})^2$);  
b) mixed SM-DFV decays, where one top will decay into $Wb$ and the other into $u_i\, \mu \, \tau$ 
(Br$(t \bar t)|_{SM-DFV}=$ $2{\mathcal{B}}(1-{\mathcal{B}})$); 
c) purely DFV decays (DFV-DFV), in this case both the top quarks will decay into $u_i\, \mu \, \tau$ 
($Br(t\bar t)|_{DFV}= \mathcal{B}^2$).
%%%
Because a purely DFV decay is strongly suppressed, we will focus only on SM-DFV top quark decays. 
As in the channel $t \bar t \to u_i\tau \mu W b$ there is no charge correlation between the $\mu$
and the $W$, if the $W$ decays leptonically ($W \rightarrow \ell \nu_\ell$) it is 
possible to have both like-sign (LS) and opposite-sign (OS) final states: $\mu^\pm \ell^\pm$ and 
$\mu^\pm \ell^\mp$.
%%%
On the other hand, the $\tau$ lepton can decay hadronically with a probability around 65.5 \%, producing a $\tau$-jet 
containing a small number of charged and neutral hadrons. When the momentum of the $\tau$ is 
large compared to the mass a very collimated jet is produced. At LHC center of mass energy, 
for a transverse momentum $P_T>$ $50$ GeV$/c$, 
90\% of the energy is contained in a cone of radius $R=\sqrt{ \Delta \eta^2 + \Delta \Phi^2}=$ $0.2$. 
Hadronic $\tau$ decays have low charged track multiplicity (one or three prongs) and a relevant 
fraction of electromagnetic energy deposition in the calorimeters due to photons coming from
the decay of neutral pions. In Table 1 we show the expected $\sigma \times$BR for 
the LS $t \bar t$ decays: $t \bar t \rightarrow \mu^\pm \ell^\pm u_i b \tau_h \nu_\ell$.
%%%
\begin{table}[t!]
\begin{center}
\begin{tabular}{|c|c|c|c|c|c|}
\hline $m_{H}($GeV/$c^2)$ & $114.4$ & $130$ & $150$ & $160$ & $170$ \\
\hline
Cross section [fb]          &$122.$ & $97.$ & $46.$ &  $20.$ &  $3.$\\
\hline
\end{tabular}
\caption{Cross section times the branching ratio for the LS $t \bar t$ decay: 
$t \bar t \rightarrow \mu^\pm \ell^\pm u_i b \tau_h \nu_\ell$
($\ell=$ $e$, $\mu$).} \label{tbl1}
\end{center}
\vspace{-0.9cm}
\end{table}
%%%
%%%
We then look for $t \bar t$ final states, containing a tau lepton decaying hadronically 
($\tau_h$) and two LS leptons: $t \bar t \rightarrow \mu^\pm \ell^\pm u_i b \tau_h \nu_\ell$
($\ell=$ $e$, $\mu$).
%%%
%%%
Events are selected requiring the presence of at least two jets (including the $b$-jet), 
a $\mu$, another lepton ($e$ or $\mu$ with the same sign of the first $\mu$), 
one tagged $\tau$-jet, one tagged $b$-jet and missing transverse energy.
%%
%%% LEPTONS
The two LS leptons are required to have $p_T>$ $20$ GeV/$c^2$ and 
$\eta<$ $2.4$.
%%%  JETS
Jets are reconstructed with a cone size of $R=\sqrt{(\Delta \eta)^2 + (\Delta \phi)^2} = 0.5$,
a trasverse energy of $E^{\rm jet}_T > 40~{\rm GeV}$ within a fiducial volume of 
$|\eta| <2.4$.
%%%  B-jet
The b-jets are tagged and required to pass the same kinematic cuts as
those applied on the jets.
%%%  TAU
The presence of $\tau$, in the events, is guaranteed by tagging the $\tau$-jets\cite{ATLASCMS}.  
A further cut on the transverse energy of the $\rm \tau$-jet, is applied in order to 
further suppress the background. 
Using the first $\rm 30~fb^{-1}$ of data, it will be possible to 
exclude, at 95 \% C.L., DFV top quark decays mediated by a higgs 
with a mass up to 160 GeV/$c^2$.

%%%%%%%%%%%%%%%%%%%%%%%%%%%%%%%%%%%%%%%%%%%%%%%%%%%%%%%%%%%%%%%%%%%%%%%%%%%%%%%
\section{Conclusions}
%%%%%%%%%%%%%%%%%%%%%%%%%%%%%%%%%%%%%%%%%%%%%%%%%%%%%%%%%%%%%%%%%%%%%%%%%%%%%%%

\label{c}In this paper we studied the possibility of detecting double
flavor violating top quark decays $t \to u_i\,\mu\,\tau$ 
in a model--independent manner using the effective Lagrangian approach.
A Yukawa sector that includes all the $SU_L(2)\times U_Y(1)$
invariant operators of up to dimension six was proposed and used
to construct the most general renormalizable flavor violating
$q_iq_jH$ and $Hl_il_j$ vertices. Low--energy data were used to
constraint the $tu_iH$ and $H\tau \mu$ couplings and then used to
predict the double flavor violating decay $t\to u_i\tau \mu$. It
was found that this decay has a branching ratio of order 
of $ 10^{-3}-10^{-4}$ for $115$ GeV$/c^2$$<m_H<2m_W$. Considering that
LHC will operate as a veritable top quark factory, by using the first
$\rm 30~fb^{-1}$ of data it will be possible to exclude, at 95\% C.L., 
DFV top quark decays up to a higgs mass of 160 GeV/$c^2$.
%%%%%%%%%%%%%%%%%%%%%%%%%%%%%%%%%%%%%%%%%%%%%%%%%%%%%%%%%%%%%%%%%%%%%%%%%%%%%%%

%%%%%%%%%%%%%%%%%%%%%%%%%%%%%%%%%%%%%%%%%%%%%%%%%%%%%%%%%%%%%%%%%%%%%%%%%%%%%%%
\section{Acknowledgments}
We thank the University of Cassino (Italy), CONACYT (Mexico) and the HELEN ALFA-EC program 
for their financial support.
%%%%%%%%%%%%%%%%%%%%%%%%%%%%%%%%%%%%%%%%%%%%%%%%%%%%%%%%%%%%%%%%%%%%%%%%%%%%%%%
%
%
%%%%%%%%%%%%%%%%%%%%%%%%%%%%%%%%%%%%%%%%%%%%%%%%%%%%%%%%%%%%%%%%%%%%%%%%%%%%%%%

%%%%%%%%%%%%%%%%%%%%%%%%%%%%%%%%%%%%%%%%%%%%%%%%%%%%%%%%%%%%%%%%%%%%%%%%%%%%%%%
%
\end{document}